\documentstyle[preprint,aps,eqsecnum,tighten]{revtex}
\begin{document}
\preprint{UAHEP 976}
\draft
\title{Perturbations in the Kerr-Newman Dilatonic
Black Hole Background \\
I. Maxwell Waves}
\author{R. Casadio, B. Harms and Y. Leblanc}
\address{Department of Physics and Astronomy,
The University of Alabama\\
Box 870324, Tuscaloosa, AL 35487-0324}
\author{P.H. Cox}
\address{Physics Department, Texas A\&M University-Kingsville\\
Kingsville, TX 78363}
\maketitle
\begin{abstract}
In this paper we analyze the perturbations of the Kerr-Newman
dilatonic black hole background.
For this purpose we perfom a double expansion in both the background
electric charge and the wave parameters of the relevant quantities
in the Newman-Penrose formalism.
We then display the gravitational, dilatonic and electromagnetic
equations, which reproduce the static solution (at zero order in the
wave parameter) and the corresponding wave equations in the Kerr
background (at first order in the wave parameter and zero order in the
electric charge).
At higher orders in the electric charge one encounters corrections
to the propagation of waves induced by the presence of a non-vanishing
dilaton.
An explicit computation is carried out for the electromagnetic waves
up to the asymptotic form of the Maxwell field perturbations
produced by the interaction with dilatonic waves.
A simple physical model is proposed which could make these perturbation
relevant in the detection of radiation coming from the region of space
near a black hole.
\end{abstract}
\section{Introduction}
\label{intro}
The theory of extended objects such as strings and membranes
is a very attractive candidate for the quantum theory of gravity.
The drawback of the theory of extended objects is
that the theory is anomaly free only in some space whose dimension
is much larger than one in which we can make measurements.
This drawback is partially mitigated by the fact that a
four-dimensional, low energy effective action can be written
in terms of the fields associated with the massless excitations
of extended objects.
In a series of papers \cite{hl1}-\cite{hl7} we have explored
the viewpoint that quantum black holes are massive excitations
of extended objects and hence are elementary particles.
In Ref.\cite{chlc}, starting from the low energy effective action
describing the Einstein-Maxwell theory interacting with a dilaton
in four dimensions $(G = 1)$,
\begin{eqnarray}
S = {1\over16\,\pi}\,\int d^4x\,\sqrt{-g}\,\left[R-{1\over2}\,
(\nabla\phi)^2-e^{-a\,\phi}\,F^2\right]
\ ,
\label{action}
\end{eqnarray}
we obtained the static solutions of the field equations
for a Kerr-Newman dilaton black hole rotating with arbitrary
angular momentum by expanding the fields in terms of the
charge-to-mass ratio of the black hole.
In the present paper we use these solutions, which appear as
source terms in the wave equations, to determine the effect
of a scalar component of gravity (the dilaton) on the
propagation of electromagnetic waves in the vicinity of a charged,
rotating black hole.
\par
Our goal is to obtain expressions which can be compared to
experimentally measurable quantities in order to test the idea that
quantum black holes are extended objects.
In all likelihood the scalar component of gravity will have a
coupling to electromagnetic waves which is as weak as that of
the tensor component.
Therefore the best place to look for the effect of the dilaton on
these waves is in the neighborhood of a cosmological black hole, which
we consider to be composed of quantum black holes.
The expressions we obtain in this paper show how the electromagnetic
radiation reaching a distant observer from such black holes is affected
by the presence of a scalar component of gravity.
\par
We begin in Section~\ref{stat} with a summary of the static background
solutions for a Kerr-Newman dilatonic black hole.
Perturbative expressions for the metric tensor components and the
electromagnetic fields are given through order $Q^2$.
Expressions for the gravitational field, the dilaton field and
the electromagnetic field in terms of the Newman-Penrose
formalism complete this section.
\par
In Section~\ref{formal} we obtain the wave equations for the various
field modes.
Our method is to double expand each field in powers of the electric
charge and the wave parameters.
Substituting these expansions into Maxwell's equations, the
dilaton equation and Einstein's equations then give (inhomogeneous) wave
equations for the coefficients of the expansion to any desired accuracy.
Part of the currents in the wave equations arise due to the presence of
the dilaton.
We give explicit expressions for the Maxwell wave equations for the
coefficients linear in the wave parameter for the three lowest orders
of the charge parameter.
\par
The forms of the solutions of the wave equations at large $r$ are
examined in Section~\ref{asy}.
We derive explicit expressions for the corrections to the Maxwell
scalars for both the ingoing and outgoing waves.
An infrared problem which arises in the amplitudes of these corrections
is also discussed.
In the same Section we analyze the corrections to the ingoing and
outgoing waves near the horizon.
\par
In Section~\ref{det} we discuss the production of outgoing electromagnetic
waves by the dilatonic waves which occur in the case of dynamic
black holes.
We give the explicit contributions to the electric and magnetic fields
arising from the solutions we found in the previous Section and
also use these contributions to write the expression for the
correction to the energy flux due to the presence of a dilaton field.
\par
Section~\ref{close} is a discussion of our results and of the further
investigations these results allow us to pursue.
\section{The static background solution}
\label{stat}
The action describing the Einstein-Maxwell theory interacting
with a dilaton in four dimensions is given in Eq.~(\ref{action}),
where $R$ is the scalar curvature,
$\phi$ is the dilaton field, $a$ is the dilaton parameter and
$F$ is the Maxwell field.
In tensor components, $i,j=0,\ldots,3$,
the corresponding field equations are
\begin{eqnarray}
\left\{\begin{array}{l}
\nabla^i(e^{-a\,\phi}\,F_{ij}) = 0  \\
\nabla_{[ k}F_{ij]} = 0 \; , \\
\end{array}\right.
 \ \ {\rm (Maxwell)}
\label{maxwell}
\\
 \nonumber \\
\nabla^2\phi = -a\,e^{-a\,\phi}\,F^2 \ \ \ \ {\rm (dilaton)}\; ,
\label{phi}
\\
\nonumber \\
R_{ij} = {1\over{2}}\,\nabla_i\phi\,\nabla_j\phi + 2\,
T^{EM}_{ij} \ \
{\rm (Einstein)}
\ ,
\label{einstein}
\end{eqnarray}
where $\nabla$ is the covariant derivative,
$R_{ij}$ is the Ricci tensor, and
the electromagnetic energy--momentum tensor is given as
\begin{eqnarray}
T^{EM}_{ij} = e^{-a\,\phi}\,\left[ F_{ik}\,F^k_j-{1\over{4}}\,g_{ij}\,
F^2\right]
\ .
\end{eqnarray}
\par
In Ref.~\cite{chlc} we found a perturbative static solution of the field
equations above representing a charged, rotating black hole of the
Kerr-Newman type with a non-vanishing background dilaton field
in which the parameter of expansion is the charge-to-mass
ratio, $Q/M$ \cite{stima}.
\subsection{Perturbative $Q/M$ solution}
The metric in our solution is of the general form which describes
a stationary axisymmetric space-time \cite{chan}
\begin{eqnarray}
ds^2 = -{\Delta\,\sin^2\theta\over\Psi}\,(dt)^2
+ \Psi\,(d\varphi - \omega\,dt)^2
+\rho^2\,\left[{(dr)^2\over\Delta}+(d\theta)^2\right]
\; ,
\label{g_ij}
\end{eqnarray}
in which $x^0=t$, $x^1=r$, $x^2=\theta$, $x^3=\varphi$.
\par
The explicit expressions of the functions $\Psi=\Psi(r,\theta)$,
$\omega=\omega(r,\theta)$ and $\Delta=\Delta(r)$
as displayed in Ref.~\cite{chlc} can be simplified upon substituting
for the (bare) parameters $M$, $Q$ and $J\equiv \alpha\,M$
the quantities
\begin{eqnarray}
M_{phys} &\equiv& M\, \left[ 1 + {a^2\,Q^2\over{6\,M^2}}\right]
\nonumber \\
Q_{phys} &\equiv& Q\,\left[1 + {a^2\,Q^2\over{3\,M^2}}\right]
\nonumber \\
J_{phys} &\equiv&\alpha\,M_{phys}
\ ,
\end{eqnarray}
which are respectively the ADM mass, charge and angular momentum
of the hole which determine the geodesic motion in the
asymptotically flat region.
We also introduce the following shift in the radial coordinate
\begin{eqnarray}
r\to r-{a^2\,Q^2_{phys}\over6\,M_{phys}}
\ ,
\end{eqnarray}
and we finally get that the metric in Eq.~(\ref{g_ij})
coincides (at order $Q^2_{phys}/M^2_{phys}$) with the
pure Kerr-Newman solution, that is
\begin{eqnarray}
\Delta&=&r^2-2\,M_{phys}\,r+\alpha^2_{phys}+Q_{phys}^2
\nonumber \\
\rho^2&=&r^2+\alpha^2_{phys}\,\cos^2\theta
\nonumber \\
\Psi&=&-{\Delta-\alpha^2_{phys}\,\sin^2\theta\over\rho^2}
\nonumber \\
\omega&=&-\alpha_{phys}\,\sin^2\theta\,[1+\Psi^{-1}]
\ ,
\label{KN-metric}
\end{eqnarray}
where we have also written $\alpha_{phys}=\alpha$.
This also implies that the singularity structure is not affected by
the presence of the dilaton at that order.  In the general
case however (if one includes the higher order corrections in $Q^2/M^2$)
this is not so, because the spacetime is very likely
no longer a Petrov type-D spacetime.
In the present case one still has two horizons for
$\Delta=0$, that is
\begin{equation}
r_\pm=M_{phys} \pm\sqrt{M^2_{phys}-\alpha^2_{phys}- Q^2_{phys}}
\ .
\end{equation}
\par
However the presence of a non-zero dilaton field
\begin{eqnarray}
\phi=-a\,{r\over\rho^2}\,{Q_{phys}^2\over M_{phys}}
\ ,
\label{p0}
\end{eqnarray}
affects the electric and magnetic field potentials $A$ and $B$
(see Ref.~\cite{chlc} for the definitions),
\begin{eqnarray}
A&=&Q_{phys}\,{r\over\rho^2}\,\left[
1-\left({1\over2\, r}+{r\over\rho^2}\right)\,
{a^2\,Q_{phys}^2\over 3\,M_{phys}}\right]
\nonumber \\
B&=&-Q_{phys}\,\alpha_{phys}\,{\cos\theta\over\rho^2}\,
\left[1-\left({1\over 2\,M}-{r\over\rho^2}\right)\,
{a^2\,Q^2_{phys}\over3\,M_{phys}}\right]
\ ,
\end{eqnarray}
where the terms proportional to $Q^2_{phys}$ inside
the brackets correspond to the corrections with respect to
the Kerr-Newman potentials \cite{chan}.
These corrections also affect the electric and magnetic fields
in the limit $r\to\infty$,
\begin{eqnarray}
{\cal E}_{\hat{r}} &\approx& {Q_{phys}\over{r^2}}
\nonumber \\
{\cal E}_{\hat{\theta}} &\approx&
-{2\,\alpha^2_{phys}\, Q_{phys}\over{r^4}}\,\sin\theta\,\cos\theta
\nonumber \\
{\cal B}_{\hat{r}} &\approx&
{2\,\alpha_{phys}\, Q_{phys}\over{r^3}}\,\cos\theta
\,\left[1 - {a^2\,Q^2_{phys}\over{6\,M^2_{phys}}}\right]
\nonumber \\
{\cal B}_{\hat{\theta}} &\approx&
{\alpha_{phys}\, Q_{phys}\over{r^3}}\,
\sin\theta\left[1 - {a^2\,Q^2_{phys}\over{6\,M^2_{phys}}}\right]
\ ,
\end{eqnarray}
where ${\cal E}_{\hat a}$ and ${\cal B}_{\hat a}$ ($\hat a
=\hat r, \hat\theta$)
are respectively
the electric and the magnetic field components with respect to the
usual spherical coordinate tetrad basis \cite{misner}.
One would thus experience a relative shift in the intensity of
the field ${\cal B}$ with respect to ${\cal E}$.
This is also the source for the anomalous gyromagnetic ratio
\begin{eqnarray}
g=2\,\left[1-{a^2\,Q_{phys}^2\over 6\, M^2_{phys}}\right]
\ ,
\end{eqnarray}
which should be compared with $g=2$ for the Kerr-Newman black hole
\cite{strau}.
\par
For the sake of simplicity, from now on we omit the subscript $phys$
in all $Q$, $M$ and $\alpha$.
\subsection{Newman-Penrose form of the solution}
In the next Section we will compute the linear perturbation
field equations for the gravitational field, Eq.~(\ref{einstein}),
the scalar dilaton field, Eq.~(\ref{phi}), and the Maxwell field,
Eq.~(\ref{maxwell}), in this background.
We will work in the Newman-Penrose formalism \cite{NP,chan},
which is a special type of tetrad calculus in which one introduces
four null vectors $e_{a}^i$, $a=1,\ldots,4$,
\begin{eqnarray}
l^i&\equiv&e_{1}^i
\nonumber \\
n^i&\equiv&e_{2}^i
\nonumber \\
m^i&\equiv&e_{3}^i
\nonumber \\
m^{* i}&\equiv&e_{4}^i
\ ,
\end{eqnarray}
at every point of space-time, where $m^{\ast i}$ is the complex
conjugate of $m^i$.
They must satisfy the normalization conditions
\begin{eqnarray}
l^i\,n_i=1\ \ &{\rm and}&\ \ m^i\,m^*_i=-1
\ .
\end{eqnarray}
Further $l^i$ is affinely parameterized while the other vectors
are not.
\par
Since our static metric coincides with the Kerr-Newman one,
and the latter differs from the Kerr metric only in the definition
of $\Delta$ \cite{chan}, all the quantities formally computed in the
Kerr case apply.
In detail, one has the following expressions for the four vectors
\begin{eqnarray}
l^i&=&{1\over\Delta}\,\left(r^2+\alpha^2,+\Delta,0,\alpha\right)
\nonumber \\
n^i&=&{1\over2\,\rho^2}\,\left(r^2+\alpha^2,-\Delta,0,\alpha\right)
\nonumber \\
m^i&=&{1\over\sqrt{2}\,\bar\rho}\,
\left(i\,\alpha\,\sin\theta,0,1,i\,\csc\theta\right)
\ ,
\end{eqnarray}
where
\begin{eqnarray}
\bar\rho\equiv r+i\,\alpha\,\cos\theta\ \ \ \ &{\rm and}&
\ \ \ \
\bar\rho^*\equiv r-i\,\alpha\,\cos\theta
\ .
\end{eqnarray}
Also, the directional derivatives along the four null vectors are given
special symbols,
\begin{eqnarray}
\hat D&\equiv& l^i\,\partial_i
\nonumber \\
\hat\Delta&\equiv& n^i\,\partial_i
\nonumber \\
\hat\delta&\equiv& m^i\,\partial_i
\ .
\end{eqnarray}
\par
In the tetrad formalism covariant derivatives (with respect to
a coordinate basis) are replaced by intrinsic derivatives
(see Ref.~\cite{chan}) represented by the symbol $|$.
For scalar quantities one has $f_{|c}=f_{,c}$, for covariant
vectors
\begin{eqnarray}
V_{a|c}\equiv V_{a,c}-\eta^{bd}\,\gamma_{bac}\,V_d
\ ,
\end{eqnarray}
and so forth, where the indices $a,b,\ldots=1,\ldots,4$ are
tetrad indices, $\eta^{bd}={\rm diag}(-1,+1,+1,+1)$ is the Minkowski
tensor and $\gamma_{bac}$ are the rotation or
{\em spin coefficients}.
For the Kerr-Newman metric they are represented by the following
Greek letters and have the expressions \cite{chan}
\begin{eqnarray}
&\kappa=\sigma=\lambda=\nu=\epsilon=0&
\nonumber \\
&\tilde\rho=-{1\over\bar\rho^*}\ ,\ \ \
\beta={\cot\theta\over 2\,\sqrt{2}\,\bar\rho}\ ,\ \ \
\pi={i\,\alpha\,\sin\theta\over\sqrt{2}\,(\bar\rho^*)^2}\ ,\ \ \
\tau=-{i\,\alpha\,\sin\theta\over\sqrt{2}\,\bar\rho^2}&
\nonumber \\
&\mu=-{\Delta\over2\,\rho^2\,\bar\rho^*}\ ,\ \ \
\gamma=\mu+{r-M\over2\,\rho^2}\ ,\ \ \
\tilde\alpha=\pi-\beta^*
\ .&
\label{KN-spin}
\end{eqnarray}
Further, among the Weyl scalars $\Psi_k$, $k=0,\ldots,4$,
there is only one non vanishing quantity, that is
\begin{eqnarray}
\Psi_2\equiv C_{ijkl}\,l^i\,m^j\,n^k\,m^{*l}
=-{M\over(\bar\rho^*)^3}+
{Q^2\over\rho^2\,(\bar\rho^*)^2}
\ .
\end{eqnarray}
We can thus easily claim that our solution is still of Petrov
type D up to ${\cal O}(Q^2)$ since the Kerr-Newman one is of this type.
\par
The Maxwell field can be fully described by the following three
complex quantities
\begin{eqnarray}
\phi_0&=&F_{ij}\,l^i\,m^j
\nonumber \\
\phi_1&=&{1\over 2}\,F_{ij}\,(l^i\,n^j+m^i\,m^{*j})
\nonumber \\
\phi_2&=&F_{ij}\,m^{*i}\,n^j
\ ,
\end{eqnarray}
which contain all the information about the six components
of the electric and magnetic fields.
For the background solution, they are shown to satisfy
the {\em phantom gauge} \cite{chan} at order $Q^2$,
\begin{eqnarray}
\phi_0&=&\phi_2={\cal O}(Q^3)
\ ,
\label{phantom}
\end{eqnarray}
with the third scalar given by
\begin{eqnarray}
\phi_1&=&-i\,{Q\over2\,(\bar\rho^*)^2}+{\cal O}(Q^3)
\ .
\label{max00}
\end{eqnarray}
\section{Formal development of the wave equations}
\label{formal}
We want now to study the linear perturbations of the static solution
given in the previous Section and obtain the wave equations for the
various field modes.
The full set of equations one obtains in the Newman-Penrose formalism
for a metric of the form given in Eq.~(\ref{g_ij}) is extraordinarily
large, and we refer to the book by Chandrasekhar~\cite{chan} for the
details.
In Ref.~\cite{chan} it is shown that, while in the Kerr background the
Maxwell and gravitational wave equations decouple,
in the Kerr-Newman background (even without the dilaton field) they
do not.
\par
Since our solution at order $Q^2$ coincides with the Kerr-Newman one
as far as the metric is concerned, we cannot thus expect the dilaton,
Maxwell and gravitational wave equations to disentangle.
However, at order $Q^0$ it is just the Kerr solution, and we can
take advantage of the related simplifications.
For this purpose, we perform a double expansion of every relevant
quantity,
\begin{eqnarray}
&&\phi(t,r,\theta,\varphi)=\sum\limits_{p,n}\,g^p\,Q^n\,\phi^{(p,n)}
\nonumber \\
&&\phi_i(t,r,\theta,\varphi)=\sum\limits_{p,n}\,g^p\,Q^n\,
\phi^{(p,n)}_i\ ,
\ \ \ i=0,1,2
\nonumber \\
&&G(t,r,\theta,\varphi)=\sum\limits_{p,n}\,g^p\,Q^n\,G^{(p,n)}
\ ,
\label{(l,n)}
\end{eqnarray}
where $g$ is the same wave parameter for the dilaton, Maxwell and
gravitational field quantities, the latter (including tetrad components)
being collectively represented with the symbol $G$.
Although we have formally included every order in the wave parameter
in the series above, we will only study the linear ($p=1$) case.
\par
For $p=0$ the static solutions are
\begin{eqnarray}
\phi^{(0,n)}&=&\phi^{(0,n)}(r,\theta)
\nonumber \\
\phi_i^{(0,n)}&=&\phi_i^{(0,n)}(r,\theta)\ ,\ \ \ i=0,1,2
\nonumber \\
G^{(0,n)}&=&G^{(0,n)}(r,\theta)
\ .
\end{eqnarray}
At order $(0,0)$ we recover the Kerr solution with neither Maxwell
nor dilaton fields.
The metric tensor elements for this solution are obtained from
those defined in Eq.~(\ref{KN-metric}) by setting $Q=0$.
At order $(0,1)$ the Maxwell fields appear, corresponding to
the metric of the Kerr-Newman solution which is obtained at
order $(0,2)$ together with the first non-vanishing contribution
to the dilaton background.
Finally our solution as given in \cite{chlc}
is fully displayed at order $(0,3)$.
\par
Correspondingly, one expects to obtain at order $(1,0)$ the already
known, decoupled and separable wave equations in the Kerr background.
At order $(1,1)$ new contributions relating to the presence of the
dilaton waves are expected.
At order $(1,2)$ the effect induced by the dilaton background
should appear.
In the following we will show that this is actually the case.
\par
We shall also assume that the perturbations to which the various
quantities are subject have the following time and azimuthal dependence
\begin{eqnarray}
\phi^{(1,n)}(t,r,\theta,\varphi)&=&k_d\,
e^{i\,\bar\omega\,t+i\,m\,\varphi}\,\phi^{(1,n)}(r,\theta)
\nonumber \\
\phi_i^{(1,n)}(t,r,\theta,\varphi)&=&k_{EM}\,
e^{i\,\bar\omega\,t+i\,m\,\varphi}\,
\phi_i^{(1,n)}(r,\theta)\ ,\ \ \ i=0,1,2
\nonumber \\
G^{(1,n)}(t,r,\theta,\varphi)&=&k_G\,
e^{i\,\bar\omega\,t+i\,m\,\varphi}\,
G^{(1,n)}(r,\theta)
\ ,
\end{eqnarray}
where $k_d$, $k_{EM}$ and $k_G$ are (possibly equal) parameters.
We note here that each function of $r$ and $\theta$ on the R.H.S.s
above implicitly carries an extra integer index, $m$, and the continuous
dependence on the frequency $\bar\omega$ (both can be positive or
negative).
\par
The directional derivatives when acting on the wave modes displayed
above can be written
\begin{eqnarray}
\hat D&=&{\cal D}_0
\nonumber \\
\hat \Delta&=&-{\Delta\over2\,\rho^2}\,{\cal D}_0^\dagger
\nonumber \\
\hat \delta&=&{1\over\sqrt{2}\,\bar\rho}\,{\cal L}_0^\dagger
\nonumber \\
\hat \delta^*&=&{1\over\sqrt{2}\,\bar\rho^*}\,{\cal L}_0
\ ,
\end{eqnarray}
where
\begin{eqnarray}
{\cal D}_n&=&\partial_r+i\,{K\over\Delta}+2\,n\,{r-M\over\Delta}
\nonumber \\
{\cal D}_n^\dagger&=&\partial_r-i\,{K\over\Delta}+2\,n\,{r-M\over\Delta}
\nonumber \\
{\cal L}_n&=&\partial_\theta+\tilde Q+n\,\cot\theta
\nonumber \\
{\cal L}_n^\dagger&=&\partial_\theta-\tilde Q+n\,\cot\theta
\ ,
\label{cal D}
\end{eqnarray}
with $n$ an integer such that $n\ge 0$ and
\begin{eqnarray}
K&\equiv&(r^2+\alpha^2)\,\bar\omega+\alpha\,m
\nonumber \\
\tilde Q&\equiv&\alpha\,\bar\omega\,\sin\theta+m\,\csc\theta
\ .
\end{eqnarray}
\subsection{Gravitational equations}
\label{G_w}
Following Ref.~\cite{teukolski}, we consider the following
three non vacuum equations,
\begin{eqnarray}
(\hat\delta^*-4\,\tilde\alpha+\pi)\,\Psi_0
-(\hat D-4\,\tilde\rho-2\,\epsilon)\,\Psi_1
-3\,\kappa\,\Psi_2
=(\hat\delta+\pi^*-2\,\tilde\alpha^*-2\,\beta)\,R_{11}
\nonumber \\
-(\hat D-2\,\epsilon-2\,\tilde\rho^*)\,R_{12}
+2\,\sigma\,\,R_{21}
-2\,\kappa\,\,R_{22}
-\kappa^*\,\,R_{13}
\nonumber \\
\nonumber \\
(\hat\Delta-4\,\gamma+\mu)\,\Psi_0
-(\hat\delta-4\,\tau-2\,\beta)\,\Psi_1
-3\,\sigma\,\Psi_2
=(\hat\delta+2\,\pi^*-2\,\beta)\,R_{12}
\nonumber \\
-(\hat D-2\,\epsilon+2\,\epsilon^*-\tilde\rho^*)\,R_{13}
-\lambda^*\,\,R_{11}
-2\,\sigma\,\,R_{22}
-2\,\kappa\,\,R_{23}
\nonumber \\
\nonumber \\
(\hat D-\tilde\rho-\tilde\rho^*-3\,\epsilon+\epsilon^*)\,\sigma
-(\hat\delta-\tau+\pi^*-\tilde\alpha^*-3\,\beta)\,\kappa
-\Psi_0=0
\ .
\label{grav}
\end{eqnarray}
The Ricci tensor terms are given by the Einstein field equations,
Eq.~(\ref{einstein}),
\begin{eqnarray}
R_{ab}\equiv R_{ij}\,e^i_a\,e^j_b=
{1\over{2}}\,\phi_{|a}\,\phi_{|b} + 2\,T^{EM}_{ab}
\ .
\label{ricci}
\end{eqnarray}
Now we expand the equations above according to the expression
in Eq.~(\ref{(l,n)}).
\par
For $p=0$ we recover the static solution described in the previous
Section.
In particular one has
\begin{eqnarray}
&R_{ab}^{(0,0)}=R_{ab}^{(0,1)}=0&
\nonumber \\
&\Psi_k^{(0,0)}=\Psi_k^{(0,1)}=\Psi_2^{(0,1)}=0\ ,\ \ \ k=0,1,3,4&
\nonumber \\
&\kappa^{(0,0)}=\sigma^{(0,0)}=\nu^{(0,0)}=\lambda^{(0,0)}=
\epsilon^{(0,0)}=0&
\nonumber \\
&\kappa^{(0,1)}=\sigma^{(0,1)}=\nu^{(0,1)}=\lambda^{(0,1)}=
\epsilon^{(0,1)}=0
\ ,&
\label{G00}
\end{eqnarray}
which will be used to simplify the expressions at order $p=1$.
\par
We can express the (wave) perturbation
\begin{eqnarray}
(e^i_a)^{(1)}\equiv\sum\limits_n\,Q^n\,(e^i_a)^{(1,n)}
\ ,
\end{eqnarray}
in the tetrad vectors as a linear combination of the unperturbed
basis vectors $(e^i_a)^{(0,n)}$ \cite{chan},
\begin{eqnarray}
(e^i_a)^{(1,n)}=\sum\limits_{q+r=n}\,(A^b_a)^{(q)}\,(e^i_a)^{(0,r)}
\ .
\end{eqnarray}
The perturbations in the basis vectors are then fully described by
the matrix
\begin{eqnarray}
A^b_a=\sum\limits_n\,Q^n\,(A^b_a)^{(n)}
\ ,
\end{eqnarray}
whose elements $A^1_1$, $A^2_2$, $A^1_2$ and
$A^2_1$ are real, while all the others are complex
(complex conjugate to each other when changing any index 3 with
4 or viceversa).
It then follows that $A^b_a$ has sixteen independent components,
and it is also useful to define
\begin{eqnarray}
&&
\begin{array}{llll}
F^1_2\equiv{\Delta\over2\,\rho^2}\,A^1_2\ ,\ \ \ &
F^2_1\equiv{2\,\rho^2\over\Delta}\,A^2_1\ ,\ \ \ &
F^3_1\equiv{1\over\bar\rho^*}\,A^3_1\ ,\ \ \ &
F^4_1\equiv{1\over\bar\rho}\,A^4_1
 \\
F^1_3\equiv{\Delta\over2\,\rho^2\,\bar\rho}\,A^1_3\ ,\ \ \ &
F^2_3\equiv{1\over\bar\rho}\,A^2_3\ ,\ \ \ &
F^3_2\equiv{\Delta\over2\,\rho^2\,\bar\rho^*}\,A^3_2\ ,\ \ \ &
F^4_2\equiv{\Delta\over2\,\rho^2\,\bar\rho}\,A^4_2
 \\
F^1_4\equiv{\Delta\over2\,\rho^2\,\bar\rho^*}\,A^1_4\ ,\ \ \ &
F^2_4\equiv{1\over\bar\rho^*}\,A^2_4\ ,\ \ \ &
F^3_4\equiv{1\over(\bar\rho^*)^2}\,A^3_4\ ,\ \ \ &
F^4_3\equiv{1\over(\bar\rho)^2}\,A^4_3
\ ,
\end{array}
\label{Fab}
\end{eqnarray}
and
\begin{eqnarray}
\begin{array}{ll}
F=F^1_3+F^1_4\ ,\ \ \ &
B_1=F^3_1+F^3_2+F^4_1+F^4_2
\\
G=F^2_3+F^2_4\ ,\ \ \ &
B_2=F^3_1+F^3_2-F^4_1-F^4_2
\\
H=F^1_3+F^3_2\ ,\ \ \ &
C_1=F^1_3+F^2_3-F^1_4-F^2_4
\\
J=F^4_1-F^4_2\ ,\ \ \ &
C_2=F^1_3-F^2_3-F^1_4+F^2_4
\\
U=A^1_1+A^2_2\ ,\ \ \ &
V=A^3_3+A^4_4
\ .
\end{array}
\label{F}
\end{eqnarray}
We are now ready to study the perturbations with $p=1$.
\subsubsection{$(1,0)$ order}
It is easy to see that
\begin{eqnarray}
R_{ab}^{(1,0)}=0
\ ,
\end{eqnarray}
so that both the electromagnetic field and the dilaton field
decouple from the gravitational field equations.
Further, on also using Eq.~(\ref{G00}), the system of gravitational
equations reduces to
\begin{eqnarray}
&&
(\hat\delta^*-4\,\tilde\alpha+\pi)^{(0,0)}\,\Psi_0^{(1,0)}
-(\hat D-4\,\tilde\rho)^{(0,0)}\,\Psi_1^{(1,0)}
-3\,\kappa^{(1,0)}\,\Psi_2^{(0,0)}=0
\nonumber \\
&&
(\hat\Delta-4\,\gamma+\mu)^{(0,0)}\,\Psi_0^{(1,0)}
-(\hat\delta-4\,\tau-2\,\beta)^{(0,0)}\,\Psi_1^{(1,0)}
-3\,\sigma^{(1,0)}\,\Psi_2^{(0,0)}=0
\nonumber \\
&&
(\hat D-\tilde\rho-\tilde\rho^*)^{(0,0)}\,\sigma^{(1,0)}
-(\hat\delta-\tau+\pi^*-\tilde\alpha^*-3\,\beta)^{(0,0)}\,
\kappa^{(1,0)}-\Psi_0^{(1,0)}=0
\ ,
\label{G10}
\end{eqnarray}
which are the equations for gravitational waves in the Kerr background
\cite{teukolski}.
\subsubsection{$(1,n\ge 1)$ order}
From the expression of the Ricci tensor, Eq.~(\ref{ricci}), one obtains
that the background dilaton field $\phi^{(0,2)}$ enters into the
equations for the gravitational waves only at order $Q^n$ with $n\ge 2$.
\par
In fact, at order $(1,1)$ one has
\begin{eqnarray}
R_{ab}^{(1,1)}&=&2\,T_{ab}^{EM\,(1,1)}
\nonumber \\
&=&
2\,F_{ik}^{(1,0)}\,F^{k\,(0,1)}_{\ j}
+2\,F_{ik}^{(0,1)}\,F^{k\,(1,0)}_{\ j}
\nonumber \\
&&-{1\over 2}\,g_{ij}^{(0,0)}\,\left(F_{kl}^{(1,0)}\,F^{kl\,(0,1)}
+F_{kl}^{(0,1)}\,F^{kl\,(1,0)}\right)
\ .
\end{eqnarray}
Thus the dilaton field does not affect the equations at this order
in the $Q$ expansion, neither directly nor through its effect on
the Maxwell waves ($F_{ij}^{(1,0)}$ are free electromagnetic waves
in Kerr background, see Section~\ref{max_w} below).
However, both Maxwell background and wave fields already enter into the
gravitational equations, so that for $n\ge 1$ it is no longer possible
to decouple gravitational waves from Maxwell excitations.
\par
At order $(1,2)$ one obtains terms containing products of the form
$\phi^{(0,2)}_{|a}\,\phi^{(1,0)}_{|b}$ which couple the dilaton
waves to their own background.
Further, one expects corrections coming from effects induced on the
electromagnetic waves.
For this reason and because of the complex entanglement among all kinds
of waves, we do not address this issue in the present paper.
\par
We just notice here that it is only at order $(1,4)$ that purely
background terms like $\phi^{(0,2)}_{|a}\,\phi^{(0,2)}_{|b}$ appear.
However, our knowledge of the static solution does not allow us
to write explicitly the equations in Eq.~(\ref{grav}) at this order,
since we do not know the gravitational terms $G^{(0,4)}$.
\subsection{Dilaton equation}
\label{dila_w}
The equation for the dilaton field, Eq.~(\ref{phi}), in tetrad
components becomes
\begin{eqnarray}
\eta^{ab}\,\phi_{a|b}=-a\,e^{-a\,\phi}\,F^2
\ ,
\label{dila2}
\end{eqnarray}
where $\phi_a\equiv e^i_a\,\phi_{,i}$ are four (complex) scalars.
\par
In the Newman-Penrose tetrad, Eq.~(\ref{dila2}) becomes
\begin{eqnarray}
&&\left[\hat D\,\hat\Delta+\hat\Delta\,\hat D
-\hat\delta\,\hat\delta^*-\hat\delta^*\,\hat\delta
+(\epsilon+\epsilon^*-\tilde\rho-\tilde\rho^*)\,\hat\Delta
+(\mu+\mu^*-\gamma-\gamma^*)\,\hat D\right.
\nonumber \\
&&\left.\phantom{[}
+(\tau-\pi^*+\tilde\alpha^*-\beta)\,\hat\delta^*
+(\tau^*-\pi+\tilde\alpha-\beta^*)\,\hat\delta\right]\phi=
-a\,e^{-a\,\phi}\,F^2
\ ,
\label{dila}
\end{eqnarray}
which can now be expanded in $Q$ and the wave parameter.
\par
At order $(0,n)$, $n=0,1,2$, one recovers the static solutions
\begin{eqnarray}
\phi^{(0,0)}=\phi^{(0,1)}=0
\nonumber \\
\phi^{(0,2)}=-{a\over M}\,{r\over\rho^2}
\ ,
\label{dila00}
\end{eqnarray}
and we now study the linear perturbations $p=1$.
\subsubsection{$(1,0)$ order}
The R.H.S. of Eq.~(\ref{dila}) vanishes and one finds the
Klein-Gordon equation for a free scalar field in Kerr space-time,
\begin{eqnarray}
&&\left[\hat D\,\hat\Delta+\hat\Delta\,\hat D
-\hat\delta\,\hat\delta^*-\hat\delta^*\,\hat\delta
-(\tilde\rho+\tilde\rho^*)\,\hat\Delta
+(\mu+\mu^*-\gamma-\gamma^*)\,\hat D\right.
\nonumber \\
&&\left.\phantom{[}
+(\tau-\pi^*+\tilde\alpha^*-\beta)\,\hat\delta^*
+(\tau^*-\pi+\tilde\alpha-\beta^*)\,\hat\delta\right]^{(0,0)}
\phi^{(1,0)}=0
\ .
\label{dila10}
\end{eqnarray}
This can be further simplified to the final form
\begin{eqnarray}
\left[\Delta_0\,{\cal D}_1\,{\cal D}_0^\dagger
+{\cal L}_0^\dagger\,{\cal L}_1
+2\,i\,\bar\omega\,\bar\rho\right]\Phi=0
\ ,
\end{eqnarray}
with $\Phi\equiv\phi^{(1,0)}$, and the operators inside the square
brackets are to be evaluated at order $(0,0)$.
\par
The equation above is separable.
One can write
\begin{eqnarray}
\Phi(r,\theta)=R_0(r)\,S_0(\theta)
\ ,
\end{eqnarray}
and then refer to the general case for (half) integer spin $s$
functions $R_s(r)\,S_s(\theta)$ described in Section~\ref{max_w}.
\subsubsection{$(1,n\ge 1)$ order}
As for the gravitational waves, the dilaton background $\phi^{(0,2)}$
enters into Eq.~(\ref{dila}) only at order $Q^2$ and higher.
\par
At order $(1,1)$ the R.H.S. of Eq.~(\ref{dila}) is different from zero
and couples the scalar field waves to the Maxwell field,
\begin{eqnarray}
\left[\Delta_0\,{\cal D}_1\,{\cal D}_0^\dagger
+{\cal L}_0^\dagger\,{\cal L}_1
+2\,i\,\bar\omega\,\bar\rho\right]\phi^{(1,1)}&=&
-a\,{k_{EM}\over k_d}\,
\left(F_{ab}^{(1,0)}\,F^{ab\,(0,1)}
+F_{ab}^{(0,1)}\,F^{ab\,(1,0)}\right)
\nonumber \\
&=&-4\,a\,{k_{EM}\over k_d}\,
\left(\phi_1^{(1,0)}\,\phi_1^{(0,1)}-
\phi_1^{*\,(1,0)}\,\phi_1^{*\,(0,1)}\right)
\ ,
\label{dila11}
\end{eqnarray}
where use has been made of the phantom gauge, Eq.~(\ref{phantom}),
and $\phi_1^{(1,0)}$ is the Maxwell wave in the Kerr background, which
we will review in the next Subsection.
Here we only observe that Eq.~(\ref{dila11}) does not contain
a contribution from the static dilaton background, neither explicitly
nor implicitly in the metric and Maxwell fields.
\par
At order $(1,2)$ the L.H.S. of Eq.~(\ref{dila}) contains terms of the
form $G^{(1,0)}\,\phi^{(0,2)}$ which couple the gravitational waves to
the dilaton background as claimed.
\subsection{Maxwell equations}
\label{max_w}
The second equation for the Maxwell field in Eq.~(\ref{maxwell})
is automatically satisfied once we introduce the three complex
scalars $\phi_i$, and we are then left with only the first equation
which can be written in a general tetrad frame as
\begin{eqnarray}
F_{ab|c}=a\,\phi_{|c}\,F_{ab}
\ .
\label{max2}
\end{eqnarray}
Thus one clearly sees that the dilaton generates a source term for
the electromagnetic field.
\par
Using the symbols for directional derivatives and spin coefficients
introduced in the previous Section, Eq.~(\ref{max2}) can be written as
the following set of four complex equations
\begin{eqnarray}
&&(\hat D-2\,\tilde\rho)\,\phi_1
-(\hat\delta^*+\pi-2\,\tilde\alpha)\,\phi_0
+\kappa\,\phi_2=J_{1}
\nonumber \\
&&(\hat\delta-2\,\tau)\,\phi_1
-(\hat\Delta+\mu-2\,\gamma)\,\phi_0
+\sigma\,\phi_2=J_{3}
\nonumber \\
&&(\hat D-\tilde\rho+2\,\epsilon)\,\phi_2
-(\hat\delta^*+2\,\pi)\,\phi_1
+\lambda\,\phi_0=J_{4}
\nonumber \\
&&(\hat\delta-\tau+2\,\beta)\,\phi_2
-(\hat\Delta+2\,\mu)\,\phi_1
+\nu\,\phi_0=J_{2}
\ ,
\label{max}
\end{eqnarray}
where the source terms are given by
\begin{eqnarray}
J_{1}&=&{a\over 2}\,\left[(\phi_1+\phi_1^*)\,\hat D
-\phi_0\,\hat\delta^*-\phi_0^*\,\hat\delta\right]\phi
\nonumber \\
J_{2}&=&{a\over 2}\,\left[(\phi_1+\phi_1^*)\,\hat\Delta
-\phi_2\,\hat\delta-\phi_2^*\,\hat\delta^*\right]\phi
\nonumber \\
J_{3}&=&{a\over 2}\,\left[(\phi_1-\phi_1^*)\,\hat\delta
-\phi_0\,\hat\Delta+\phi_2^*\,\hat D\right]\phi
\nonumber \\
J_{4}&=&{a\over 2}\,\left[(\phi_1-\phi_1^*)\,\hat\delta^*
+\phi_0^*\,\hat\Delta-\phi_2\,\hat D\right]\phi
\ .
\end{eqnarray}
We can now expand according to Eq.~(\ref{(l,n)}).
\subsubsection{$(1,0)$ order}
We observe that the background dilaton field and the
Maxwell background field are zero at order $Q^0$.
This makes the source terms vanish and, using the
vanishing of some of the spin coefficients according to
Eq.~(\ref{G00}), one obtains
\begin{eqnarray}
&&
(\hat D-2\,\tilde\rho)^{(0,0)}\,\phi_1^{(1,0)}
-(\hat\delta^*+\pi-2\,\tilde\alpha)^{(0,0)}\,\phi_0^{(1,0)}=0
\nonumber \\
&&
(\hat\delta-2\,\tau)^{(0,0)}\,\phi_1^{(1,0)}
-(\hat\Delta+\mu-2\,\gamma)^{(0,0)}\,\phi_0^{(1,0)}=0
\nonumber \\
&&
(\hat D-\tilde\rho)^{(0,0)}\,\phi_2^{(1,0)}
-(\hat\delta^*+2\,\pi)^{(0,0)}\,\phi_1^{(1,0)}=0
\nonumber \\
&&
(\hat\delta-\tau+2\,\beta)^{(0,0)}\,\phi_2^{(1,0)}
-(\hat\Delta+2\,\mu)^{(0,0)}\,\phi_1^{(1,0)}=0
\ .
\label{max10}
\end{eqnarray}
These are the equations for electromagnetic waves $\phi_i^{(1,0)}$
moving in the Kerr background.
Their properties and solutions have already been broadly studied
in the literature and we now briefly summarize them for the purpose
of introducing some useful notation \cite{chan}.
\par
We replace the directional derivatives and spin coefficients
with their explicit forms as given respectively in Eq.~(\ref{cal D})
and Eq.~(\ref{KN-spin}) and obtain
\begin{eqnarray}
&&
\left({\cal L}_1-{i\,\alpha\,\sin\theta\over\bar\rho^*}\right)
\Phi_0
-\left({\cal D}_0+{1\over\bar\rho^*}\right)\Phi_1=0
\nonumber \\
&&
\left({\cal L}_0+{\,i\,\alpha\,\sin\theta\over\bar\rho^*}\right)
\Phi_1
-\left({\cal D}_0-{1\over\bar\rho^*}\right)\Phi_2=0
\nonumber \\
&&
\left({\cal L}_1^\dagger-{i\,\alpha\,\sin\theta\over\bar\rho^*}
\right)\Phi_2
+\Delta_0\,\left({\cal D}_0^\dagger+{1\over\bar\rho^*}\right)
\Phi_1=0
\nonumber \\
&&
\left({\cal L}_0^\dagger+{\,i\,\alpha\,\sin\theta\over\bar\rho^*}
\right)\Phi_1
+\Delta_0\,\left({\cal D}_1^\dagger-{1\over\bar\rho^*}\right)
\Phi_0=0
\ ,
\end{eqnarray}
where $\Delta_0\equiv\Delta^{(0,0)}$, and we have also introduced
\begin{eqnarray}
\Phi_0&=&\phi_0^{(1,0)}
\nonumber \\
\Phi_1&=&\sqrt{2}\,{\bar\rho^*}\,\phi_1^{(1,0)}
\nonumber \\
\Phi_2&=&2\,(\bar\rho^*)^2\,\phi_2^{(1,0)}
\ .
\end{eqnarray}
On using the following commutation properties
\begin{eqnarray}
&&\left[\left({\cal D}_0+{1\over\bar\rho^*}\right),
\left({\cal L}_1^\dagger+{i\,\alpha\,\sin\theta\over\bar\rho^*}\right)
\right]=0
\nonumber \\
&&\left[\Delta_0\,\left({\cal D}_0^\dagger+{1\over\bar\rho^*}\right),
\left({\cal L}_0+{i\,\alpha\,\sin\theta\over\bar\rho^*}\right)
\right]=0
\ ,
\end{eqnarray}
one can obtain separate equations for $\Phi_0$ and $\Phi_2$,
\begin{eqnarray}
&&
\left[\Delta_0\,{\cal D}_0\,{\cal D}_0^\dagger
+{\cal L}_0^\dagger\,{\cal L}_1
-2\,i\,\bar\omega\,\bar\rho\right]\Delta_0\,\Phi_0=0
\nonumber \\
&&
\left[\Delta_0\,{\cal D}_0^\dagger\,{\cal D}_0
+{\cal L}_0\,{\cal L}_1^\dagger
+2\,i\,\bar\omega\,\bar\rho\right]\Phi_2=0
\ .
\end{eqnarray}
The latter are separable equations and we can factorize
their solutions
\begin{eqnarray}
&&\Delta_0\,\Phi_0=P_{+1}(r)\,S_{+1}(\theta)\ ,
\ \ \ P_{+1}=\Delta_0\,R_{+1}
\nonumber \\
&&\Phi_2=P_{-1}(r)\,S_{-1}(\theta)\ ,
\ \ \ P_{-1}=R_{-1}
\ .
\end{eqnarray}
The radial equations for the functions $P_{s=\pm1}$ are particular
cases of the eigenvalue equations,
\begin{eqnarray}
&&
\left[\Delta_0\,{\cal D}_{1-|s|}\,{\cal D}_0^\dagger
-2\,i\,(2\,|s|-1)\,\bar\omega\,r\right]P_{|s|}=E\,P_{|s|}
\nonumber \\
&&
\left[\Delta_0\,{\cal D}_{1-|s|}^\dagger\,{\cal D}_0
+2\,i\,(2\,|s|-1)\,\bar\omega\,r\right]P_{-|s|}=E\,P_{-|s|}
\ ,
\label{radials}
\end{eqnarray}
where $P_{|s|}\equiv\Delta_0^s\,R_s$ and $P_{-|s|}\equiv R_{-s}$,
with $s$ integer or half integer.
The angular equations for $S_{\pm1}$ are particular cases
of the following equations
\begin{eqnarray}
&&
\left[{\cal L}_0^\dagger\,{\cal L}_{|s|}
+2\,\bar\omega\,\alpha\,\cos\theta \right]S_{|s|}=-E\,S_{|s|}
\nonumber \\
&&
\left[{\cal L}_0\,{\cal L}_{|s|}^\dagger
-2\,\bar\omega\,\alpha\,\cos\theta\right]S_{-|s|}=-E\,S_{-|s|}
\ ,
\label{angulars}
\end{eqnarray}
where $E$ is the same (separation) constant which appears
in Eq.~(\ref{radials}).
\par
For any (half) integers $|s|\le l$ and $|m|\le 2\,l+1$ one has
\begin{eqnarray}
S^m_{sl}(\theta,\bar\omega)\,e^{i\,m\,\varphi}=
Y^m_{sl}(\theta,\varphi;\bar\omega)
\ ,
\end{eqnarray}
where $Y(\bar\omega)$ are {\em spin-weighted
spheroidal harmonics\/}
\cite{X} which form a complete, orthonormal set of functions for every
value of $s$.
They reduce to the {\em spin-weighted spherical harmonics\/},
\begin{eqnarray}
Y^m_{sl}(\theta,\varphi)=S^m_{sl}(\theta)\,e^{i\,m\,\varphi}
\ ,
\end{eqnarray}
in the limit $\bar\omega=0$ and to the usual spherical harmonics
when one has also $s=0$.
\par
Since the spin-weighted spherical harmonics of a fixed spin $s$
are complete and orthonormal as well, one can perform the
following expansion
\begin{eqnarray}
S^m_{sl}(\theta;\bar\omega)=\sum\limits_{l'}\,
A^m_{ss',ll'}(\bar\omega)\,S^m_{s'l'}(\theta)
\ ,
\end{eqnarray}
where the $A^m_{ss',ll'}$ are related to Clebsch-Gordan coefficients.
\par
We now introduce the following expansions for the Maxwell
wave modes $\Phi_i=\Phi_i^m(r,\theta;\bar\omega)$ of energy
$\bar\omega$ and angular momentum $m$ along the axis
of symmetry,
\begin{eqnarray}
\Phi_i(r,\theta)&=&\sum\limits_{l\ge(|m|-1)/2}\,
C_{sl}^m(\bar\omega)\,R^m_{sl}(r,\bar\omega)\,
S^m_{sl}(\theta,\bar\omega)
\nonumber \\
&=&\sum\limits_{l',l\ge(|m|-1)/2}\,B^m_{ss',ll'}(\bar\omega)\,
R^m_{sl}(r,\bar\omega)\,S^m_{s'l'}(\theta)
\ ,
\end{eqnarray}
where $s=+1$ for $i=0$ and $s=-1$ for $i=2$, the functions
$R(r,\bar\omega)$ are solutions of the radial equations,
Eq.~(\ref{radials}), $C_{sl}^m(\bar\omega)$ are normalization
constants and
\begin{eqnarray}
B^m_{ss',ll'}(\bar\omega)\equiv
C_{sl}^m(\bar\omega)\,A^m_{ss',ll'}(\bar\omega)
\ ,
\end{eqnarray}
are related to Clebsch-Gordan coefficients and project the angular
solutions on the complete basis of a generic spin $s'$.
By analogy, the dilaton wave $\Phi$ can be expanded as
\begin{eqnarray}
\Phi(r,\theta)&=&\sum\limits_{l\ge(|m|-1)/2}\,
C_{0l}^m(\bar\omega)\,R^m_{0l}(r,\bar\omega)\,
S^m_{0l}(\theta,\bar\omega)
\nonumber \\
&=&\sum\limits_{l',l\ge(|m|-1)/2}\,B^m_{00,ll'}(\bar\omega)\,
R^m_{0l}(r,\bar\omega)\,S^m_{l'}(\theta)
\ ,
\end{eqnarray}
where $S^m_l\,e^{i\,m\,\varphi}$ are spherical harmonics.
\subsubsection{$(1,1)$ order}
Since for the background solution one has
$G^{(0,1)}=\phi_0^{(0,1)}=\phi_1^{(0,1)}=0$,
the first equation in Eq.~(\ref{max}) can be written,
\begin{eqnarray}
&k_{EM}\,\left[(\hat D-2\,\tilde\rho)^{(0,0)}\,\phi_1^{(1,1)}
-(\hat\delta^*+\pi-2\,\tilde\alpha)^{(0,0)}\,\phi_0^{(1,1)}\right]
+k_{G}\,(\hat D-2\,\tilde\rho)^{(1,0)}\,\phi_1^{(0,1)}&
\nonumber \\
&={a\over 2}\,k_{d}\,(\phi_1+\phi_1^*)^{(0,1)}\,\hat D^{(0,0)}\,
\phi^{(1,0)}
\ ,&
\end{eqnarray}
which contains contributions from both
the dilaton waves $\phi^{(1,0)}$ and the gravitational waves
$G^{(1,0)}$.
\par
However, it is possible to show that the gravitational waves
actually cancel out in the Kerr background \cite{chan}.
In fact, one has
\begin{eqnarray}
(\hat D-2\,\tilde\rho)^{(1,0)}&=&
-(F^1_2)^{(0)}\,\left(\partial_r+{2\over\bar\rho^*}\right)
+{\sqrt{2}\,\rho^2\over\Delta_0}\,F^{(0)}\,
\left(\partial_\theta+{2\,i\,\alpha\,\sin\theta\over\bar\rho^*}\right)
\nonumber \\
(\hat\delta-2\,\tau)^{(1,0)}&=&
-\bar\rho^*\,H^{(0)}\,\left(\partial_r+{2\over\bar\rho^*}\right)
+{\bar\rho^*\over\sqrt{2}}\,(F^3_4)^{(0)}\left(\partial_\theta+
{2\,i\,\alpha\,\sin\theta\over\bar\rho^*}\right)
\nonumber \\
(\hat\delta^*+2\,\pi)^{(1,0)}&=&
-\bar\rho\,J^{(0)}\,\left(\partial_r+{2\over\bar\rho^*}\right)
+{\bar\rho\over\sqrt{2}}\,(F^4_3)^{(0)}\left(\partial_\theta+
{2\,i\,\alpha\,\sin\theta\over\bar\rho^*}\right)
\nonumber \\
(\hat\Delta+2\,\mu)^{(1,0)}&=&
-{\Delta_0\over 2\,\rho^2}\,(F^2_1)^{(0)}\,
\left(\partial_r+{2\over\bar\rho^*}\right)
+{1\over\sqrt{2}}\,G^{(0)}\,\left(\partial_\theta+
{2\,i\,\alpha\,\sin\theta\over\bar\rho^*}\right)
\ ,
\end{eqnarray}
where the functions $F^a_b$, $F$, $H$, $G$ are defined in
Eqs.~(\ref{Fab}), (\ref{F}).
Further, from Eq.~(\ref{max00}), one finds
\begin{eqnarray}
\left(\partial_r+{2\over\bar\rho^*}\right)\,\phi_1^{(0,1)}=
\left(\partial_\theta+{2\,i\,\alpha\,\sin\theta\over\bar\rho^*}\right)
\,\phi_1^{(0,1)}=0
\ ,
\end{eqnarray}
so that all the gravitational wave terms vanish.
\par
One is then left with the following four equations
\begin{eqnarray}
&&(\hat D-2\,\tilde\rho)\,\phi_1^{(1,1)}
-(\hat\delta^*+\pi-2\,\tilde\alpha)\,\phi_0^{(1,1)}
=a\,{k_{d}\over k_{EM}}\,\phi^R_1\,\hat D\,\Phi
\nonumber \\
&&(\hat\delta-2\,\tau)\,\phi_1^{(1,1)}
-(\hat\Delta+\mu-2\,\gamma)\,\phi_0^{(1,1)}
=a\,{k_{d}\over k_{EM}}\,\phi^I_1\,\hat \delta\,\Phi
\nonumber \\
&&(\hat D-\tilde\rho)\,\phi_2^{(1,1)}
-(\hat\delta^*+2\,\pi)\,\phi_1^{(1,1)}
=a\,{k_{d}\over k_{EM}}\,\phi^I_1\,\hat \delta^*\,\Phi
\nonumber \\
&&(\hat\delta-\tau+2\,\beta)\,\phi_2^{(1,1)}
-(\hat\Delta+2\,\mu)\,\phi_1^{(1,1)}
=a\,{k_{d}\over k_{EM}}\,\phi^R_1\,\hat \Delta\,\Phi
\ ,
\label{max11}
\end{eqnarray}
where we have defined $\phi^{R/I}_1\equiv(\phi_1\pm\phi_1^*)^{(0,1)}/2$,
$\Phi=\phi^{(1,0)}$ and all gravitational quantities
are understood to be of order $(0,0)$.
\par
The L.H.S. of Eq.~(\ref{max11}) above is the same as the L.H.S.
of Eq.~(\ref{max10}), so we introduce
\begin{eqnarray}
W_0&=&\phi_0^{(1,1)}
\nonumber \\
W_1&=&\sqrt{2}\,{\bar\rho^*}\,\phi_1^{(1,1)}
\nonumber \\
W_2&=&2\,(\bar\rho^*)^2\,\phi_2^{(1,1)}
\nonumber \\
\Phi^{I/R}_1&=&\sqrt{2}\,{\bar\rho^*}\,\phi^{I/R}_1
\ ,
\end{eqnarray}
and follow the same steps previously done at order $(1,0)$.
We then obtain separate equations for $W_0$ and $W_2$,
\begin{eqnarray}
&&
\left[
\left({\cal L}_0^\dagger+{i\,\alpha\,\sin\theta\over\bar\rho^*}\right)\,
\left({\cal L}_1-{i\,\alpha\,\sin\theta\over\bar\rho^*}\right)
+\Delta_0\,\left({\cal D}_1+{1\over\bar\rho^*}\right)\,
\left({\cal D}_1^\dagger-{1\over\bar\rho^*}\right)\right]\,W_0=
T_+
\nonumber \\
&&
\left[
\left({\cal L}_0+{i\,\alpha\,\sin\theta\over\bar\rho^*}\right)\,
\left({\cal L}_1^\dagger-{i\,\alpha\,\sin\theta\over\bar\rho^*}\right)
+\Delta_0\,\left({\cal D}_0^\dagger+{1\over\bar\rho^*}\right)\,
\left({\cal D}_0-{1\over\bar\rho^*}\right)\right]\,W_2=
\Delta_0\,T_-
\ ,
\end{eqnarray}
which can be further simplified to the final form
\begin{eqnarray}
&&
\left[\Delta_0\,{\cal D}_0\,{\cal D}_0^\dagger
+{\cal L}_0^\dagger\,{\cal L}_1
-2\,i\,\bar\omega\,\bar\rho\right]\Delta_0\,W_0
=\Delta_0\,T_+
\nonumber \\
&&
\left[\Delta_0\,{\cal D}_0^\dagger\,{\cal D}_0
+{\cal L}_0\,{\cal L}_1^\dagger
+2\,i\,\bar\omega\,\bar\rho\right] W_2
=\Delta_0\,T_-
\ .
\label{max11f}
\end{eqnarray}
We observe that the L.H.S.s of the equations above are again separable,
since they coincide with the expressions one gets at order $(1,0)$.
\par
The currents on the R.H.S.s are given by
\begin{eqnarray}
T_+&=&-a\,{k_d\over k_{EM}}\,\left[
\left({\cal D}_0+{1\over\bar\rho^*}\right)\,\Phi^I_1\,{\cal L}_0^\dagger
-\left({\cal L}_0^\dagger+{i\,\alpha\,\sin\theta\over\bar\rho^*}\right)\,
\Phi^R_1\,{\cal D}_0\right]\Phi
\nonumber \\
&=&{\sqrt{2}\over 2}\,a\,{k_d\over k_{EM}}\,\left[
\left(\partial_r+i\,{K\over\Delta_0}+{1\over\bar\rho^*}\right)\,
i\,\bar\rho^*\,\partial_r\left({r\over\rho^2}\right)\,
\left(\partial_\theta-\tilde Q\right)\right.
\nonumber \\
&&\phantom{{\sqrt{2}\over 2}\,a\,{k_d\over k_{EM}}\,[}\left.
-\left(\partial_\theta-\tilde Q+{i\,\alpha\,\sin\theta\over\bar\rho^*}
\right)\,{\bar\rho^*\over\alpha\,\sin\theta}\,\partial_\theta
\left({r\over\rho^2}\right)\,
\left(\partial_r+i\,{K\over\Delta_0}\right)\right]\Phi
\nonumber \\
T_-&=&a\,{k_d\over k_{EM}}\,\left[
\left({\cal L}_0+{i\,\alpha\,\sin\theta\over\bar\rho^*}\right)\,
\Phi^R_1\,{\cal D}_0^\dagger
-\left({\cal D}_0^\dagger+{1\over\bar\rho^*}\right)\,
\Phi^I_1\,{\cal L}_0\right]\Phi
\nonumber \\
&=&{\sqrt{2}\over 2}\,a\,{k_d\over k_{EM}}\,\left[
\left(\partial_r-i\,{K\over\Delta_0}+{1\over\bar\rho^*}\right)\,
i\,\bar\rho^*\,\partial_r\left({r\over\rho^2}\right)\,
\left(\partial_\theta+\tilde Q\right)\right.
\nonumber \\
&&\phantom{{\sqrt{2}\over 2}\,a\,{k_d\over k_{EM}}\,[}\left.
-\left(\partial_\theta+\tilde Q+{i\,\alpha\,\sin\theta\over\bar\rho^*}
\right)\,{\bar\rho^*\over\alpha\,\sin\theta}\,\partial_\theta
\left({r\over\rho^2}\right)\,
\left(\partial_r-i\,{K\over\Delta_0}\right)\right]\Phi
\ ,
\end{eqnarray}
and they cannot be written as the sums of functions of $r$ only
and $\theta$ only.
\par
However one can expand the currents on the angular basis functions
we have introduced at order $(1,0)$,
\begin{eqnarray}
J_{\pm,l}^m(r,\bar\omega)&=&\Delta_0(r)\,\int_{-1}^{+1} d(\cos\theta)\,
T_\pm(r,\theta)\,S^m_{\pm 1,l}(\theta,\bar\omega)
\nonumber \\
&=&
\Delta_0(r)\,\sum\limits_{l'}\,\int_{-1}^{+1} d(\cos\theta)\,
T_\pm(r,\theta)\,B_{0,ll'}^m(\bar\omega)\,
S^m_{l'}(\theta)
\ ,
\end{eqnarray}
and define
\begin{eqnarray}
W_{+,l}^m(r,\bar\omega)&=&\Delta_0(r)\,\int_{-1}^{+1} d(\cos\theta)\,
W_0(r,\theta)\,S^m_{+1,l}(\theta,\bar\omega)
\nonumber \\
W_{-,l}^m(r,\bar\omega)&=&\int_{-1}^{+1} d(\cos\theta)\,
W_2(r,\theta)\,S^m_{-1,l}(\theta,\bar\omega)
\ .
\end{eqnarray}
One then obtains the following radial equations for the
functions $W_\pm$,
\begin{eqnarray}
&&
\left[\Delta_0\,{\cal D}_0\,{\cal D}_0^\dagger
-2\,i\,\bar\omega\,r-E_{+1,l}^m\right] W_{+,l}^m
=J_{+,l}^m
\nonumber \\
&&
\left[\Delta_0\,{\cal D}_0^\dagger\,{\cal D}_0
+2\,i\,\bar\omega\,r-E_{-1,l}^m\right] W_{-,l}^m
=J_{-,l}^m
\ .
\label{max11c}
\end{eqnarray}
Given the explicit form of the currents, analytic solutions
of these equations do not appear to be feasible.
\par
In this Section we only observe that, without the dilaton ($a=0$),
one would not expect any dependence on the charge $Q$ in the form
of the Maxwell waves at order $(1,1)$, that is we can assume
\begin{eqnarray}
\phi_i^{(1,1)}(a=0)=0
\ .
\end{eqnarray}
But $\phi_i^{(1,1)}(a=0)$ are also the solutions of the homogeneous
equations derived by setting $T_\pm$ to zero  in Eq.~(\ref{max11f}).
It thus follows that the particular solutions $W_0$ and $W_2$ of the
inhomogeneous Eq.~(\ref{max11f}) precisely represent a purely dilatonic
effect.
\subsubsection{$(1,2)$ order}
So far, the dilaton background field has not appeared in our
Maxwell equations.
However, one expects $\phi^{(0,2)}$ to enter in the expressions of the
currents in the R.H.S.s of Eq.~(\ref{max}) at order $(1,2)$ and in fact
it does.
Unfortunately, the L.H.S.s of the same equations now contain both
gravitational and electromagnetic waves, which are known to be coupled
in the Kerr-Newman background \cite{chan}.
\par
In order to proceed, one can make the following working {\em ansatz\/}:
we assume
\begin{eqnarray}
k_{EM}\gg k_d,k_G
\ ,
\label{kkk}
\end{eqnarray}
and neglect both dilaton and gravitational waves.
This hypothesis is equivalent to assuming (space-time) boundary
conditions
such that the gravitational and dilaton contents of the wave field
are negligibly small when compared to the electromagnetic sources.
\par
We also observe that the currents $T_\pm$ at order $(1,1)$ are
proportional to $k_d/k_{EM}$ and become negligible when Eq.~(\ref{kkk})
is satisfied.
Thus the corresponding particular solutions $\phi_i^{(1,1)}$ for the
inhomogeneous Maxwell equations vanish as well.
On the other hand, when Eq.~(\ref{kkk}) does not apply
($k_d\sim k_{EM}$), one can trust the solutions we found at order
$(1,1)$ and neglect the present order of approximation ($Q^2$).
This simple argument shows that the two cases, order $(1,1)$
with $k_d\sim k_{EM}$ and order $(1,2)$ with Eq.~(\ref{kkk}),
are sufficient to cover most of the electromagnetic physics one
expects to be affected by the existence of a dilaton field.
\par
The four Maxwell equations with Eq.~(\ref{kkk}) then read
\begin{eqnarray}
&&
(\hat D-2\,\tilde\rho)^{(0,0)}\,\phi_1^{(1,2)}
-(\hat\delta^*+\pi-2\,\tilde\alpha)^{(0,0)}\,\phi_0^{(1,2)}
=J_{1}^{(1,2)}
\nonumber \\
&&
(\hat\delta-2\,\tau)^{(0,0)}\,\phi_1^{(1,2)}
-(\hat\Delta+\mu-2\,\gamma)^{(0,0)}\,\phi_0^{(1,2)}
=J_{3}^{(1,2)}
\nonumber \\
&&
(\hat D-\tilde\rho)^{(0,0)}\,\phi_2^{(1,2)}
-(\hat\delta^*+2\,\pi)^{(0,0)}\,\phi_1^{(1,2)}
=J_{4}^{(1,2)}
\nonumber \\
&&
(\hat\delta-\tau+2\,\beta)^{(0,0)}\,\phi_2^{(1,2)}
-(\hat\Delta+2\,\mu)^{(0,0)}\,\phi_1^{(1,2)}
=J_{2}^{(1,2)}
\ .
\label{max12}
\end{eqnarray}
The currents on the R.H.S.s are given by
\begin{eqnarray}
J_{1}^{(1,2)} &=&
-i\,\left({K\over\Delta}\right)^{(0,2)}\,\phi_1^{(1,0)}
+{a\over 2}\,\left[(\phi_1+\phi_1^*)^{(1,0)}\,\partial_r
-(\phi_0+\phi_0^*)^{(1,0)}\,\partial_\theta
\right]\,\phi^{(0,2)}
\nonumber \\
J_{2}^{(1,2)} &=&
\left(i\,{K\over\Delta}+2\,\mu\right)^{(0,2)}\,\phi_1^{(1,0)}
+{a\over 2}\,\left[(\phi_1+\phi_1^*)^{(1,0)}\,\partial_\theta
+(\phi_0^*-\phi_2)^{(1,0)}\,\partial_r
\right]\,\phi^{(0,2)}
\nonumber \\
J_{3}^{(1,2)} &=&
\left(i\,{K\over\Delta}+3\,\mu\right)^{(0,2)}\,\phi_0^{(1,0)}
+{a\over 2}\,\left[(\phi_1-\phi_1^*)^{(1,0)}\,\partial_r
-(\phi_2+\phi_2^*)^{(1,0)}\,\partial_\theta
\right]\,\phi^{(0,2)}
\nonumber \\
J_{4}^{(1,2)} &=&
-i\,\left({K\over\Delta}\right)^{(0,2)}\,\phi_2^{(1,0)}
+{a\over 2}\,\left[(\phi_1-\phi_1^*)^{(1,0)}\,\partial_\theta
-(\phi_0-\phi_2^*)^{(1,0)}\,\partial_r
\right]\,\phi^{(0,2)}
\ ,
\end{eqnarray}
from which one concludes that the perturbations $\phi_i^{(1,2)}$
couple to the dilaton background through free Maxwell waves
($\phi_i^{(1,0)}$).
\par
However, the full consistency of the argument above requires that
the same condition in Eq.~(\ref{kkk}) be applied {\em simultaneously\/}
to all wave equations at order $(1,2)$.
Thus one obtains constraints from both the gravitational and dilaton
wave equations at order $(1,2)$.
We leave this analysis to a future investigation.
\section{Asymptotic expansions}
\label{asy}
The equations we obtained in Section~\ref{max_w} for the Maxwell waves
at order $(1,1)$ look almost intractable, due to the complexity of
the source terms.
For this reason we consider the equations in the limits $r\to\infty$
and $r\sim r_+$, and study the asymptotic regimes of ingoing and
outgoing modes.
\subsection{Ingoing modes at large radius}
We first assume that the ingoing modes of the Maxwell field
$W_0^{in}$ and $W_2^{in}$ can be approximated for large $r$
as
\begin{eqnarray}
\Delta_0\,W_0^{in}&\simeq&
A_+^{in}\,Z_+^{in}\,{e^{i\,\bar\omega\,r}\over r^{n_+}}
\nonumber \\
W_2^{in}&\simeq&
A_-^{in}\,Z_-^{in}\,{e^{i\,\bar\omega\,r}\over r^{n_-}}
\ ,
\end{eqnarray}
where $A_\pm^{in}=a_\pm^{in}+i\,b_\pm^{in}$ are (complex) coefficients,
$Z_\pm^{in}=Z_\pm^{in}(\theta;m,\bar\omega)$ are angular functions
and $n_\pm$ are integers, all to be determined later.
The L.H.S.s of Maxwell equations in Eq.~(\ref{max11f}) become
\begin{eqnarray}
&&
\left[\Delta_0\,{\cal D}_0\,{\cal D}_0^\dagger
+{\cal L}_0^\dagger\,{\cal L}_1
-2\,i\,\bar\omega\,\bar\rho\right]\Delta_0\,W_0^{in}
=
A_+^{in}\,{e^{i\,\bar\omega\,r}\over r^{n_+-1}}\,
\left\{2\,\bar\omega\,\left[2\,\bar\omega\,M-i\,(1+n_+)\right]\,
Z_+^{in}+{\cal O}\left({1\over r}\right)\right\}
\nonumber \\
&&
\left[\Delta_0\,{\cal D}_0^\dagger\,{\cal D}_0
+{\cal L}_0\,{\cal L}_1^\dagger
+2\,i\,\bar\omega\,\bar\rho\right] W_2^{in}
=
A_-^{in}\,{e^{i\,\bar\omega\,r}\over r^{n_+-1}}\,
\left\{2\,\bar\omega\,\left[2\,\bar\omega\,M+i\,(1-n_-)\right]\,
Z_-^{in}+{\cal O}\left({1\over r}\right)\right\}
\ .
\label{lhs11_ir}
\end{eqnarray}
Then we use for the dilaton field the asymptotic (ingoing) form
\cite{teukolski},
\begin{eqnarray}
\Phi^{in}=\phi^{(1,0)}(r,\theta)=
F^{in}\,S\,{e^{i\,\bar\omega\,r}\over r}
\ ,
\label{P}
\end{eqnarray}
where $F^{in}$ is a real coefficient,
$S=S_l^m(\theta; \bar\omega)$ is a spheroidal wave function
(of spin zero), and we find
\begin{eqnarray}
\Delta_0\,T_+&\simeq&{\sqrt{2}\over 2}\,a\,{k_d\over k_{EM}}\,
F^{in}\,\bar\omega\,(\partial_\theta-\tilde Q)\,S\,
e^{i\,\bar\omega\,r}
\nonumber \\
\Delta_0\,T_-&\simeq&{\sqrt{2}\over 4}\,a\,{k_d\over k_{EM}}\,
F^{in}\,(i-2\,\bar\omega\,M)\,(\partial_\theta+\tilde Q)\,S\,
{e^{i\,\bar\omega\,r}\over r}
\ .
\label{T_ir}
\end{eqnarray}
On equating the R.H.S.s in Eq.~(\ref{lhs11_ir}) and the expressions
for the currents, Eq.~(\ref{T_ir}), one gets $n_+=1$ and $n_-=2$.
Further, the coefficients $A_\pm^{in}$ and the function $S_\pm^{in}$
must satisfy the following equations
\begin{eqnarray}
&&
A_+^{in}\,\left\{4\,\bar\omega\,(\omega\,M-i)\,Z_+^{in}
+{\cal O}\left({1\over r}\right)\right\}=
{\sqrt{2}\over 2}\,a\,{k_d\over k_{EM}}\,F^{in}\,
\left\{\bar\omega\,(\partial_\theta-\tilde Q)\,S+
{\cal O}\left({1\over r}\right)\right\}
\nonumber \\
&&
A_-^{in}\,\left\{2\,\bar\omega\,(2\,\omega\,M-i)\,Z_-^{in}
+{\cal O}\left({1\over r}\right)\right\}=
{\sqrt{2}\over 4}\,a\,{k_d\over k_{EM}}\,F^{in}\,\left\{
(i-2\,\bar\omega\,M)\,(\partial_\theta+\tilde Q)\,S+
{\cal O}\left({1\over r}\right)\right\}
\ .
\label{mxr}
\end{eqnarray}
Thus one obtains the solutions
\begin{eqnarray}
\left\{\begin{array}{l}
n_+=1 \\
Z_+^{in}=(\partial_\theta-\tilde Q)\,S \\
a_+^{in}=\strut\displaystyle {\sqrt{2}\over 8}\,a\,{k_d\over k_{EM}}\,
F^{in}\,{\bar\omega\,M\over 1+\bar\omega^2\,M^2} \\
b_+^{in}=\strut\displaystyle {\sqrt{2}\over 8}\,a\,{k_d\over k_{EM}}\,
F^{in}\,{1\over 1+\bar\omega^2\,M^2}
\ ,
\end{array}\right.
\label{A+}
\end{eqnarray}
and
\begin{eqnarray}
\left\{\begin{array}{l}
n_-=2  \\
Z_-^{in}=(\partial_\theta+\tilde Q)\,S   \\
a_-^{in}=-\strut\displaystyle{\sqrt{2}\over 8}\,a\,{k_d\over k_{EM}}\,
F^{in}\,{1\over \bar\omega}
\\
b_-^{in}=0
\ .
\end{array}\right.
\label{A-}
\end{eqnarray}
\par
From the form of $a_-^{in}$ one clearly sees that this asymptotic
solution for the ingoing modes introduces some singular behaviour in
the infrared region $\bar\omega\to 0$.
Actually, in the limit $\bar\omega\to 0$ the full Eqs.~(\ref{max11f})
are well behaved with respect to $\bar\omega$ and thus we expect it
to disappear after summation of all order in the large $r$ expansion.
\par
However, for the purpose of comparing the particular (dilaton dependent)
solutions $\phi_i^{(1,1)}$ to the free waves $\phi_i^{(1,0)}$
in the black hole background, it is sufficient to consider those
wavelengths which are short enough to probe the presence of the
black hole itself.
This provides us with a natural wavelength cut-off $\bar\lambda\sim r_+$
and makes the expressions in Eqs.~(\ref{A+}) and (\ref{A-}) sufficiently
well-behaved for $\bar\omega>1/\bar\lambda$.
\subsection{Outgoing modes at large radius}
The analysis of outgoing modes can be greatly simplified once one notices
that Eq.~(\ref{max11f}) can be written
\begin{equation}
{\cal G}_\pm(\bar\omega,m)\,f_\pm=
\Delta_0\,{\cal T}_\pm(\bar\omega,m)\,
\Phi
\ ,
\label{mx}
\end{equation}
where $f_+=\Delta_0\,W_0$ and $f_-=W_2$.
The differential operators ${\cal G}_\pm$ and ${\cal T}_\pm$
share the following symmetry,
\begin{eqnarray}
&&{\cal G}_+(-\bar\omega,-m)={\cal G}_-(\bar\omega,m)
\nonumber \\
&&
{\cal T}_+(-\bar\omega,-m)={\cal T}_-(\bar\omega,m)
\ .
\end{eqnarray}
On the other side, the transformation $\bar\omega\to-\bar\omega$
(together with the relabelling $F^{in}\to F^{out}$) maps {\em ingoing\/} 
modes to {\em outgoing\/} modes, and vice versa.
Thus, it is sufficient to study one set of modes and then apply the
mapping above to obtain the solutions for the other set.
\par
For the dilaton field we have the asymptotic (outgoing) form
\cite{teukolski},
\begin{eqnarray}
\Phi^{out}=F^{out}\,S\,{e^{-i\,\bar\omega\,r}\over r}
\ ,
\label{Phi_or}
\end{eqnarray}
where $S$ is a spheroidal wave function and $F^{out}$ an arbitrary
coefficient.
Since we already know the ingoing solutions from the previous
Subsection, we can take full advantage of the symmetry shown above
and write
\begin{eqnarray}
\Delta_0\,W_0^{out}&\simeq&
A_+^{out}\,Z_+^{out}\,{e^{-i\,\bar\omega\,r}\over r^2}
\nonumber \\
W_2^{out}&\simeq&
A_-^{out}\,Z_-^{out}\,{e^{-i\,\bar\omega\,r}\over r}
\ .
\label{out}
\end{eqnarray}
Since $\tilde Q(-\bar\omega,-m)=-\tilde Q(\bar\omega,m)$,
the angular parts are now given by
\begin{eqnarray}
Z_+^{out}&=&(\partial_\theta-\tilde Q)\,S
\nonumber \\
Z_-^{out}&=&(\partial_\theta+\tilde Q)\,S
\ ,
\end{eqnarray}
and the coefficients $A_\pm^{out}=a_\pm^{out}+i\,b_\pm^{out}$
are given by
\begin{eqnarray}
\left\{\begin{array}{l}
a_\pm^{out}=-a_\mp^{in}
\\
b_\pm^{out}=b_\mp^{in}
\ .
\end{array}\right.
\end{eqnarray}
These coefficients again display the same infrared behaviour
as $A_\pm^{in}$ in the leading order of the large $r$ expansion.
The discussion on a natural cut-off, as explained in the
previous Subsection, however is equally applicable for the present
outgoing modes.
\subsection{Ingoing and outgoing modes close to the horizon}
\label{r+}
From the classical analysis of wave propagation close to the
horizon, one usually obtains special boundary conditions on the
surface $r=r_+$.
Namely, only ingoing modes (as seen by any local observer)
are allowed \cite{teukolski,chan} and there are no
non-special outgoing modes available for the dilaton waves
on the horizon.
Thus one could infer that the currents $T_\pm^{out}$ are
identically zero and $W_0^{out}=W_2^{out}\equiv 0$ on the horizon.
\par
However, one can think of a dynamical situation in which
some of the parameters of the black hole are changed and
generate gravitational and dilatonic waves whose
intensity is higher near (and outside) the horizon.
The dilaton waves can result in a source for electromagnetic
perturbations which is sufficently strong at least in a
region $r\sim r_+$.
The outgoing modes then propagate freely once they exit this
region (see next Section and Eq.~(\ref{phi_211c})).
The ingoing modes instead can be detected away from the
horizon if they satisfy the superradiant condition,
$\bar\omega<m\,\omega_+$, where $\omega_+=\alpha/(2\,M\,r_+)$
is the superradiant frequency.
\par
A useful radial parameterization of the external region
($r_+<r<+\infty$) is given by the {\em turtle\/} coordinate
\cite{teukolski,chan},
\begin{eqnarray}
r_*\simeq{2\,M\,r_+\over r_+-r_-}\,\ln(r-r_+)
\ ,\ \ \  r\sim r_+
\ ,
\end{eqnarray}
We thus find it convenient to introduce the new radial variable
\begin{equation}
x\equiv r-r_+
\ ,
\end{equation}
and then to perform an expansion of Eq.~(\ref{max11f}) at leading
order in $1/x$.
\par
We have for the dilaton field the following ingoing form
\cite{teukolski},
\begin{eqnarray}
\Phi^{in}=F^{in}\,S\,e^{i\,\tilde k\,r_*}\simeq
F^{in}\,S\,x^{i\,\Gamma}
\ ,
\label{Phi_hi}
\end{eqnarray}
where $S$ is again a spheroidal wave function, and
\begin{eqnarray}
&&\tilde k=\bar\omega-m\,\omega_+
\nonumber \\
&&\Gamma={2\,M\,r_+\,\bar\omega-m\,\alpha\over r_+-r_-}
\ .
\end{eqnarray}
We assume that the Maxwell waves are given by
\begin{eqnarray}
\Delta_0\,W_0^{in}&=&Z_+^{in}\,x^{i\,\Gamma+n_+}
\nonumber \\
W_2^{in}&=&Z_-^{in}\,x^{i\,\Gamma+n_-}
\ ,
\end{eqnarray}
and we determine the integers $n_\pm$ and the complex functions
$Z_\pm^{in}$ by equating the leading orders in $1/x$ of both sides
of Eq.~(\ref{max11f}).
\par
One easily finds $n_\pm=1$, but the forms of $S_\pm^{in}$ turn
out to be inconveniently long and we write the solutions symbolically
as
\begin{eqnarray}
&&
W_0^{in}=a\,{k_d\over k_{EM}}\,\bar\omega\,A_+^{in}\,
e^{i\,\tilde k\,r_*}
\nonumber \\
&&
W_2^{in}=a\,{k_d\over k_{EM}}\,A_-^{in}\,
\Delta_0\,e^{i\,\tilde k\,r_*}
\ ,
\label{W.in.rp}
\end{eqnarray}
where $A_\pm^{in}=A_\pm^{in}(\theta;M,m,\bar\omega)$
are complex rational functions such that
$\lim_{\bar\omega\to 0}\,A_\pm^{in}\not=0$.
Thus it turns out that $W_0$ is special on the horizon \cite{teukolski},
in that a well-behaved observer on the horizon would see it
vanish identically, while $W_2$ is not.
\par
The ougoing solutions are now easily obtained.
We assume for the dilaton field the outgoing form,
\begin{eqnarray}
\Phi^{out}=F^{in}\,S\,e^{-i\,\tilde k\,r_*}
\ ,
\label{Phi_ho}
\end{eqnarray}
and employ the symmetry between ingoing and outgoing modes
described in the previous Subsection.
The latter are then given by
\begin{eqnarray}
&&
W_0^{out}=a\,{k_d\over k_{EM}}\,A_+^{in}(-\bar\omega,-m)
\,e^{-i\,\tilde k\,r_*}
\nonumber \\
&&
W_2^{out}=a\,{k_d\over k_{EM}}\,\bar\omega\,\,A_-^{in}(-\bar\omega,-m)\,
\Delta_0\,e^{-i\,\tilde k\,r_*}
\ ,
\label{phi2}
\end{eqnarray}
and again $W_0$ is special while $W_2$ is not.
\section{Wave detection}
\label{det}
The outgoing waves at $r\to\infty$ are the ones that might possibly
be detected, thus they carry most of the information one can try
to retrieve from an experimental observation.
It is thus only with these modes that we deal here.
\par
First we observe that the solutions we get in Section~\ref{asy}
are inherently related to the existence of dilaton waves
(through the coefficients $F^{out}$), which in turn can be
interpreted as fluctuations in the gravitational-electromagnetic
coupling constant (see the form of the action in Eq.~(\ref{action})).
These fluctuations are likely to be negligible in the very asymptotic
region ($r\to\infty$).
Therefore, one does not expect them to be any appreciable source of
electromagnetic waves in the region where a detector can be possibly
located.
\par
This is consistent with our analysis at large $r$ as given in
section~\ref{asy}.
In fact the free outgoing waves at order $(1,0)$ for $r \to \infty$ are
\cite{teukolski}
\begin{eqnarray}
\phi_0^{(1,0)}&\sim& {e^{-i\,\bar\omega\,r}\over r^3}
\nonumber \\
\phi_2^{(1,0)}&\sim& {e^{-i\,\bar\omega\,r}\over r}
\ ,
\label{phi_i10}
\end{eqnarray}
and the dominant contribution comes from $\phi_2^{(1,0)}$.
Our outgoing solutions fall off more rapidly,
\begin{equation}
r\,\phi_0^{(1,1)}\sim\phi_2^{(1,1)}\sim e^{-i\,\bar\omega\,r}/r^3
\ .
\end{equation}
This implies that, when the interaction between dilaton waves and 
the background Maxwell field of the black hole takes place at large $r$,
it produces electromagnetic waves which are negligible when compared 
to free waves.
\par
However, dilatonic waves might be a strong source in a region
${\cal R}\subseteq (r_+, r_d)$ just outside the horizon (see Figure),
where they could be generated by some violent dynamical process
involving changes in the mass, charge and angular momentum of the 
black hole.
For instance, one can think of the injection of new matter from the 
accreting disk or a companion star, therefore $r_d$ would be of 
the order of the typical size of such a structure.
Also the background electromagnetic field of the black hole is stronger 
near $r_+$ and the local static fields might be even stronger if one 
takes into account possible ionization and polarization of the surrounding 
matter. 
This would further enhance the production in a manner dependent
on the charge distribution inside ${\cal R}$.
\par
Any Maxwell waves which cross the region ${\cal R}$ interact 
with all the other fields before they reach $r\to\infty$.
However, since we only consider corrections at order $(1,1)$ for which
the waves $\phi_i^{(1,1)}$ decouple from the other fields 
(except the background metric) outside ${\cal R}$, one has that,
once produced at any point inside ${\cal R}$, $\phi_i^{(1,1)}$ 
simply propagate freely (outside ${\cal R}$) to $r\to\infty$.
\par
The way we formalize this model is as follows. 
Inside ${\cal R}$, we regard $\phi_i^{(1,1)}$ to be given by particular 
outgoing solutions of the inhomogeneous equations in Eq.~(\ref{max11f}) 
with boundary conditions for the dilaton waves such that the sources 
$T_\pm$ are appreciably strong. 
Then, in the region $r>r_d$, we use these solutions as initial conditions 
at $r=r_d$ for the amplitudes of the fields $\phi_i^{(1,1)}$ which 
are now required to solve the same Eq.~(\ref{max11f}) but with 
$T_\pm\simeq 0$.
When $T_\pm=0$, the equations in Eq.~(\ref{max11f}) reduce to the 
homogeneous equations given in Eq.~(\ref{max10}) at order $(1,0)$
and whose solutions are exactly the free waves with asymptotic 
behaviour displayed above, Eq.~(\ref{phi_i10}).
Thus one has
\begin{eqnarray}
\left\{\begin{array}{l}
\phi_0^{(1,1)}(r>r_d)\simeq|\phi_0^{(1,1)}(r_d)|
\,\strut\displaystyle{r_d^3\over r^3}\,e^{-i\,\bar\omega\,r}
\\
\\
\phi_2^{(1,1)}(r>r_d)\simeq|\phi_2^{(1,1)}(r_d)|
\,\strut\displaystyle{r_d\over r}\,e^{-i\,\bar\omega\,r}
\ ,
\end{array}\right.
\label{phi_211c}
\end{eqnarray}
where $\phi_i^{(1,1)}(r_d)$ are the values taken at $r=r_d$
by the solutions of the inhomogeneous equations inside ${\cal R}$
and the matching conditions at $r=r_d$ have been applied as stated
above.
It also follows that, for $r\to\infty$, $\phi_0^{(1,1)}\sim\phi_0^{(1,0)}$
is again negligible with respect to $\phi_2^{(1,0)}$, 
but $\phi_2^{(1,1)}\sim\phi_2^{(1,0)}$ is not.
\par
If $r_d$ is relatively big (compared to $r_+$) one can still
assume the dilaton inside ${\cal R}$ to be of the form given 
in Eq.~(\ref{Phi_or}) and obtains the corresponding electromagnetic 
perturbations in Eq.~(\ref{out}).
However since ${\cal R}$ is likely to extend down to $r_+$,
we found it interesting to study Maxwell equations in a region very 
close to the horizon (see Subsection~\ref{r+}) 
In that case, the relevant form of the dilatonic waves is given in
Eq.~(\ref{Phi_ho}), and $\phi_2^{(1,1)}(r_d)$ is obtained from
Eq.~(\ref{phi2}).  
Of course a complete treatment would contain contributions from both 
the approximations.
\par
For this model one can easily compute the energy flux at infinity.
In order to do this, we first observe that
\begin{eqnarray}
\phi_0&=&{\sqrt{2}\over 2}\,\left[
{\cal E}_\theta\,{r^2+\alpha^2\over\Delta\,\bar\rho}
+{\cal B}_\phi\,{1\over\bar\rho}+i\,\left(
{\cal E}_\phi\,{\bar\rho^*\over\Delta\,\sin\theta}
-{\cal B}_\theta\,{1\over\bar\rho\,\sin\theta}\right)\right]
\nonumber \\
\phi_2&=&{\sqrt{2}\over 2}\,\left[
{\cal B}_\phi\,{\Delta\over\bar\rho^*\,\rho^2}
-{\cal E}_\theta\,{r^2+\alpha^2\over\bar\rho^*\,\rho^2}+i\,\left(
{\cal E}_\phi\,{1\over\bar\rho^*\,\sin\theta}
+{\cal B}_\theta\,{\Delta\over\bar\rho^*\,\rho^2\,\sin\theta}
\right)\right]
\ ,
\end{eqnarray}
which, in the limit $r\to\infty$ and for $\phi_0^{(1,0)}$,
$\phi_2^{(1,0)}$ given in Eq.~(\ref{phi_i10}) above,
imply that the coordinate components ${\cal E}_i$, ${\cal B}_i$
are of order $r^0$ and the local components of the electromagnetic
waves are given by
\begin{eqnarray}
&&{\cal E}_{\hat\theta}^{(1,0)}=-{\cal B}_{\hat\phi}^{(1,0)}=
-{\sqrt{2}\over 2}\,(\phi_2+\phi_2^*)^{(1,0)}
\nonumber \\
&&{\cal E}_{\hat\phi}^{(1,0)}={\cal B}_{\hat\theta}^{(1,0)}=
{\sqrt{2}\over 2\,i}\,(\phi_2-\phi_2^*)^{(1,0)}
\ .
\end{eqnarray}
The (outgoing) flux per unit solid angle at infinity is thus given by
\begin{eqnarray}
{d^2 E\over dt\,d\Omega}=\lim\limits_{r\to\infty}
{r^2\over2\,\pi}\,|\phi_2^{(1,0)}|^2
\end{eqnarray}
\par
On using the solution in Eq.~(\ref{phi_211c}) one obtains
new contributions to the electric and magnetic fields,
\begin{eqnarray}
&&{\cal E}_{\hat\theta}^{(1,1)}=-{\cal B}_{\hat\phi}^{(1,1)}\simeq
-{\sqrt{2}\over 2}\,{a_-^{out}\over r_d^3}\,|Z_-^{out}|
\nonumber \\
&&{\cal E}_{\hat\phi}^{(1,1)}={\cal B}_{\hat\theta}^{(1,1)}\simeq
{\sqrt{2}\over 2}\,{b_-^{out}\over r_d^3}\,|Z^{out}_-|
\ ,\ \ \ \ \ \
(r=r_d)
\ ,
\end{eqnarray}
where we assumed the asymptotic solution in Eq.~(\ref{out}).
These imply a correction to the energy flux,
\begin{eqnarray}
{d^2 E\over dt\,d\Omega}\simeq\lim\limits_{r\to\infty}\,
{r^2\over2\,\pi}\,\left|\phi_2^{(1,0)}+e^{i\,\gamma}\,Q\,
\phi_2^{(1,1)}\right|^2
\ ,
\end{eqnarray}
where the relative phase $\gamma$ between the $(1,0)$ term
and the $(1,1)$ term has been written explicitly.
Since the phase of the solution in Eq.~(\ref{phi_211c}) is determined
only with respect to the phase of the source (dilatonic wave) and the
latter is unrelated to $\phi_2^{(1,0)}$, the phase $\gamma$ is
essentially arbitrary.
We can also assume that dilatonic waves exist with all possible
relative phases and average over $\gamma\in(0,2\,\pi)$.
One then obtains
\begin{eqnarray}
{d^2 E\over dt\,d\Omega}&\simeq&\lim\limits_{r\to\infty}\,
{r^2\over2\,\pi}\,\int_0^{2\,\pi} {d\gamma\over2\,\pi}\,
\left|\phi_2^{(1,0)}+e^{i\,\gamma}\,Q\,
\phi_2^{(1,1)}\right|^2
\nonumber \\
&\simeq& \lim\limits_{r\to\infty}\,{r^2\over2\,\pi}\,
|\phi_2^{(1,0)}|^2+
{a^2\,Q^2\over 128\,\pi^2\,r_d^4}\,
{k_d\over k_{EM}}\,{(F^{out})^2\over 1+\bar\omega^2\,M^2}
\,|Z_-^{out}|^2
\ ,
\end{eqnarray}
or an analogous expression when the solution for $\phi_2^{(1,1)}$
inside ${\cal R}$ is of the form in Eq.~(\ref{phi2}).
The main feature of the expression above is that the correction
that we get contains information about the charge of the black hole
and the angular structure of the dilatonic perturbations close
to the horizon, the latter being encoded in the angular function
$Z_-$.
Of course, the contribution we show here is for just one frequency
$\bar\omega>1/\bar\lambda$ ($\bar\lambda$ is the cut-off discussed in
Section~\ref{asy}).
The general case is a superposition of all modes.
\section{Conclusions}
\label{close}
The double perturbative expansion (Eq.(3.1)) provides a means of
calculating the electromagnetic, dilaton and gravitational fields to any
desired accuracy.
Of course, beyond the lowest orders the field equations must be solved
numerically due to the increasing complexity of the equations with
increasing order of the expansion.
Even at the lowest order we were able to find explicit expressions
for the corrections to the electromagnetic waves only in the asymptotic
limits $r \to \infty$ and $r\sim r_+$.
We also obtained an explicit form for the correction to the
energy flux due to the presence of a scalar gravitational field (the
dilaton).
Measurement of this flux could in principle be used to test the idea that
black holes have a scalar component if the flux from a source could be
measured for two cases -- one case being when the waves from the source
are far away from a black hole and the other being when the waves pass
near to the horizon of a black hole.
An occultation observation would be a suitable event to test the idea
that black hole gravitational fields have a scalar component.
\par
In this paper we have calculated the lowest order corrections to the
electromagnetic waves propagating on a black hole background possessing
a scalar gravity component.
A future project will be to determine the corresponding corrections for
the gravitational waves propagating on the same background.
The dilaton corrections to these waves first appear at order $(1,2)$.
\acknowledgments
This work was supported in part by the U.S. Department of Energy
under Grant No. DE-FG02-96ER40967.
{\bf Figure caption}
\par\noindent
The shaded region represents the interior of the black hole ($r\le r_+$).
The region between $r_+$ and $r_d$ is the one denoted by ${\cal R}$ in
the text and corresponds to the domain where sufficiently strong
dilatonic waves interact with the background electromagnetic fields to
produce Maxwell (outgoing) radiation.
For $r>r_d$ the latter propagates freely toward the observer
($r\to\infty$).

\begin{references}
\bibitem{stima}
The ratio $Q/M$ for physical particles can be estimated
using M.K.S. units once one reintroduces the conventional
values of $G$ and $c$.
One then has $Q/M=\bar Q/(\bar M\,\sqrt{4\,\pi\,\epsilon_0\,G})$
which is approximately equal to $10^{42}$ for an electron
($\bar Q\simeq 10^{-19}$ C and $\bar M\simeq 10^{-30}$ Kg)
and to $10^{-2}$ for a particle of the same charge but of
Planck mass.
This implies that our approximation ($Q/M$ small) does not apply
to most known elementary particles, but can be valid for black holes
produced during the gravitational collapse of massive stars and
also for cosmological black holes of Planck mass size.
On the other hand, the opposite approximation ($Q/M$ large)
corresponds to a naked singularity.
\end{references}
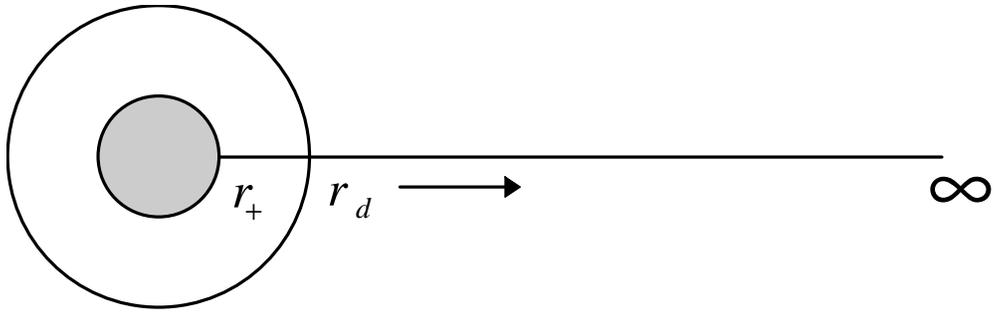
\end{document}